\documentclass[aps,preprint,nofootinbib]{revtex4}
\usepackage{framed}
\usepackage{bbold}
\usepackage{slashed}
\usepackage{graphicx,hyperref}
\usepackage{amsmath}
\usepackage{amssymb}
\usepackage{epsfig}
\usepackage{hhline}
\usepackage{soul}
\usepackage{tabularx,pifont,multirow}
\usepackage{enumitem}
\usepackage{colordvi}
\usepackage{xcolor}

\newcommand{\be}{\begin{equation}}
\newcommand{\ee}{\end{equation}}
\newcommand{\bea}{\begin{eqnarray}}
\newcommand{\eea}{\end{eqnarray}}

\newcommand{\newtext}[1]{{\textcolor{black}{#1}}}

\newcommand{\nicktext}[1]{{\textcolor{black}{#1}}}
\newcommand{\joantext}[1]{{\textcolor{black}{#1}}}

\newcommand{\nt}[1]{{\textcolor{black}{#1}}}

\newcommand{\CC}{\Lambda}
\newcommand{\rL}{\rho_{\Lambda}}
\newcommand{\rvm}{\rho^\CC_{\rm RVM}}
\newcommand{\MPl}{M_{\rm Pl}}

\newcommand{\ha}{\hat{a}}
\newcommand{\astar}{a_{*}}
\newcommand{\tHI}{\tilde{H}_I}
\newcommand{\trI}{\tilde{\rho}_I}

\newcommand{\Oro}{\Omega_{r 0}}
\newcommand{\rr}{\rho_{r}}
\newcommand{\rro}{\rho_{r 0}}
\newcommand{\rco}{\rho_{c0}}

\definecolor{darkgreen}{rgb}{0,0.3,0.05}
\makeatletter                                                         %
\newcommand*\rel@kern[1]{\kern#1\dimexpr\macc@kerna}                  %
\newcommand*\widebar[1]{                                              %
  \begingroup                                                         %
  \def\mathaccent##1##2{                                              %
    \rel@kern{0.8}                                                    %
    \overline{\rel@kern{-0.8}\macc@nucleus\rel@kern{0.2}}             %
    \rel@kern{-0.2}                                                   %
  }                                                                   %
  \macc@depth\@ne                                                     %
  \let\math@bgroup\@empty \let\math@egroup\macc@set@skewchar          %
  \mathsurround\z@ \frozen@everymath{\mathgroup\macc@group\relax}     %
  \macc@set@skewchar\relax                                            %
  \let\mathaccentV\macc@nested@a                                      %
  \macc@nested@a\relax111{#1}                                         %
  \endgroup                                                           %
}                                                                     %
\makeatother                                                          %
\begin{document}

\preprint{\leftline{KCL-PH-TH/2020-{\bf 39}}}

\title{{\bf  String-Inspired Running Vacuum, the ``Vacuumon'',  and the Swampland Criteria} \vspace{0.0cm}}

\author{Nick~E.~Mavromatos$^{a}$, \vspace{0.5cm} Joan~Sol\`a~Peracaula$^b$  and Spyros~Basilakos$^{c,d}$}

\affiliation{$^a$Theoretical Particle Physics and Cosmology Group, Physics Department, King's College London, Strand, London WC2R 2LS.
\vspace{0.5cm}\\
 $^{b}$Departament de F\'\i sica Qu\`antica i Astrof\'\i sica, \\ and \\ Institute of Cosmos Sciences (ICCUB), Univ. de Barcelona, Av. Diagonal 647 E-08028 Barcelona, Catalonia, Spain.\vspace{0.5cm}\\
$^c$Academy of Athens, Research Center for Astronomy and Applied Mathematics, Soranou Efessiou 4, 115 27 Athens, Greece. \\
$^d$ National Observatory of Athens, Lofos Nymfon,
11852, Athens, Greece.}


\begin{abstract}
\vspace{0.05cm}

We elaborate further on the compatibility of the ``vacuumon potential '' that characterises the inflationary phase of \joantext{the  Running Vacuum Model (RVM)} with the Swampland criteria. The work is motivated by the fact that, as demonstrated recently by the \nicktext{authors\,\cite{bms1,bms2,bms3}, the RVM framework} can be derived as an effective gravitational field theory stemming from underlying microscopic (critical) string theory models with gravitational anomalies,  involving condensation of primordial gravitational waves.  Although believed to be a classical \joantext{scalar field} description, not representing a fully fledged quantum field,
nonetheless we show here that the vacuumon potential satisfies certain Swampland criteria for the relevant regime of  parameters and field range. We link the criteria to the Gibbons-Hawking entropy that has been argued to characterise \joantext{the RVM} during  \joantext{the} de Sitter phase. These results imply that the vacuumon may,  after all,  admit under certain conditions, a r\^ole as a quantum field during the inflationary (almost \joantext{de}  Sitter) phase of the running vacuum. \joantext{The conventional slow-roll interpretation of this field, however,  fails just because it satisfies the Swampland criteria.
The RVM effective theory derived from the low-energy effective action of string theory  does, however, successfully describe inflation thanks to  the $\sim H^4$ terms induced by the gravitational anomalous condensates. In addition,  the stringy version of the RVM involves the Kalb-Ramond (KR) axion field, which, in contrast to the vacuumon, does perfectly satisfy the slow-roll condition. We conclude that the vacuumon description is not fully equivalent to the  stringy formulation of the RVM.  Our study provides a particularly interesting  example of  a successfully phenomenological theory beyond the $\CC$CDM,  such as the RVM,  in which the fulfilment of the  swampland criteria by the associated scalar field potential, as well as its compatibility with (an appropriate form of) the Weak-Gravity Conjecture,  prove to be insufficient conditions for warranting consistency of the scalar \nicktext{vacuum  field representation as a faithful ultraviolet complete representation of the RVM at the quantum gravity level}}.
\end{abstract}
\maketitle


\section{Introduction: Embedding Effective Field theory Models  into Quantum Gravity and the problems of cosmology} \label{sec:swamp}

Despite significant advances in models of quantum gravity, including string/brane models~\cite{gsw},
we are still far from an understanding of the microscopic theory that underlies the quantum structure of space-time at Planck scales. The string landscape~\cite{land}
\joantext{somehow adds up to the list of problems, rather than being a \nicktext{means} for resolving them}, given the enormous number of mathematically consistent string vacua, and the lack of a concrete principle to select the `physical one', other than the `anthropic principle'~\cite{anthr}, which may not be satisfactory.

On the other hand, over the past two decades,  there have been significant phenomenological advances, especially in astrophysics and Cosmology, such as the discovery of the late acceleration of the Universe, gravitational waves from coalescent black holes (via interferometric measurements), as well as `photographs' of black holes themselves~\cite{EHT}, which open up new horizons, not unrelated to quantum gravity issues.  For instance, according to the plethora of the current observational cosmological data~\cite{planck}, the observed acceleration of the Universe, can be modelled at late eras via an (approximately) de Sitter (dS) phase, in which the vacuum energy density of the Universe is dominated by a positive cosmological constant. Understanding quantum field theory (QFT) of  matter is still lacking in such background space times, even if one ignores the aspects of the quantization of the gravitational field itself. Specifically, as a result of the (observer dependent) de Sitter horizon, and thus the associated lack of an asymptotic Minkowski space-time limit, the perturbative Scattering(S)-matrix is not well defined, due to problems in defining
asymptotic quantum states. This has profound implications~\cite{Smatrix} in the inability of reconciling perturbative string theory, whose formulation is based on the existence of a well-defined S-matrix, with such dS space-times. In a similar spirit, the inflationary phase of the Universe, which also employs dS space times during the inflationary phase of the Universe, calls for a careful analysis of the associated inflaton models~\cite{inflaton}, i.e. field theory models
involving a fundamental scalar field (\emph{inflaton}), as far as their embedding in a consistent quantum gravity theory is concerned. Same holds for quintessence models~\cite{quint}, associated with relaxation models of Dark Energy via the {\it quintessence} (scalar) field.

\joantext{Apart from the above mentioned theoretical problems involving the interrelationship between cosmology, string theory, quantum gravity and effective QFT's, cosmology has to face these days also difficulties of a very practical and pedestrian nature, namely, specific problems pointing to a mismatch between theory and observations which demand urgent and efficient solutions.  \nicktext{Indeed}, in the context of the \nicktext{standard or ``concordance'' $\CC$CDM model of cosmology}, there are tensions with the present observational data --   see e.g. \cite{tensions,snowmass2020} for a review --  which must be addressed. In particular, the $\CC$CDM predicts a value of $\sigma_8$  (the mean matter fluctuations in spheres of {radius} $8h^{-1}$~Mpc, with $h$ the reduced Hubble constant)  which is in $\sim 3\sigma$ excess with respect to the direct observations, and, most significantly,  the values of the current Hubble parameter $H_0$  obtained from the local universe are $4-5\sigma$ larger  as compared to those from the CMB measurements. The persistence of these tensions strongly suggests the possibility that their explanation must be \nicktext{sought for in the framework of theoretical models}  beyond the $\CC$CDM model. This does not mean, in principle, a radical change on its structure but it may hint at the necessary presence of (\nicktext{novel, slowly evolving}) dynamical ingredients, which \nicktext{have not been taken into account}, up to now.}

\joantext{Among the existing candidate models which can mitigate these tensions, there is the ``running vacuum model'' (RVM), see \cite{JSPRev2013,JSPRev2015} for a review and references therein.  The phenomenology of this framework and its advantages as compared to $\CC$CDM in fitting the current data has been widely discussed and thoroughly analyzed  in a variety of works, see e.g. ~\cite{rvmpheno} for several of the most recent ones. It is also remarkable that frameworks mimicking the RVM (even beyond the \nicktext{General Relativity (GR)} paradigm, such as e.g. Brans-Dicke theories) \nicktext{might have} the ability to alleviate those tensions\,\cite{BD1920}. This fact is actually very important since the stamp ``effective RVM \nicktext{behaviour''} may be the kind of characteristic feature that promising models should share  in order to smooth out those tensions, as shown by the aforementioned analyses. }

\joantext{The next crucial observation for the present study is that the RVM can be derived from a four-dimensional string-inspired  low-energy effective action of graviton and antisymmetric tensor  fields of the massless (bosonic) string gravitational multiplet~\cite{bms1,bms2,bms3}.  Furthermore, and this is a point of our focus,  the RVM admits  a scalar field description, as it has recently  been put forward by the authors in Ref.\,\cite{vacuumon,rvmsugra}.  The corresponding scalar field was \nicktext{termed} the ``vacumon''. The fact that the RVM provides an efficient possible cure to the $\CC$CDM tensions}, \nicktext{combined with the existence of a scalar field representation,  naturally suggests, as a next step, to try and clarify whether the corresponding scalar field potential satisfies the swampland conjectures~\cite{dSC,SC1,SC2a,SC2b,branden}. If this question is answered in the affirmative, this would be a first step towards
embedding the vacuum model onto a consistent Ultraviolet (UV) complete quantum gravity framework. This is actually the main aim of the current  paper.  In particular, we shall concentrate on the inflationary regime of the vacuum model and check the fulfilment of the swampland criteria by the vacuumon potential in this regime. As we shall see, the swampland criteria can be satisfied under certain conditions, which, although encouraging for the embedding of the model into a more microscopic
quantum gravity framework, nonetheless implies that the vacuumon description does not satisfy the slow-roll conditions for inflation. However, from the point of view of the original four-dimensional string-inspired  version of the RVM, based on the  low-energy effective action of graviton and antisymmetric tensor  fields of the massless (bosonic) string gravitational multiplet~\cite{bms1,bms2,bms3}, condensates of graviton fluctuations provide dynamically  $\sim H^4$ contributions, which are responsible for the early de Sitter phase and the slow-roll condition is realized through the Kalb-Ramond axion field associated to the effective string action governing the inflationary regime. This implies that
the swampland criteria {\it per se}, although providing information for a possible embedding of the
effective scalar vacuumon field description of the RVM to a consistent quantum gravity framework, themselves are not {\it decisive} to yield more information on the underlying microscopic string theory features, and in particular provide any detailed information of the dynamics of the gravitational stringy condensates that underly the {\it stringy} RVM dynamics, which is itself compatible with slow-roll inflation, in the sense of admitting approximately de Sitter solutions.}

\joantext{In short, the purpose of the present work  is to analyze the scalar field representation of the string-inspired RVM in the light of the Swampland criteria  and check  the compatibility of the vacuumon picture with the original  string-inspired  version of the RVM, as well as to test if it fulfils some form of the Weak Gravity Conjectures (WGC)~\cite{WGC,Palti,ibanezWGC,kusenko}.  The layout of the paper \nicktext{is}  as follows.} \nicktext{In Sect.\,\ref{sec:Swampland}, we summarize the formulation of the various Swampland conjectures existing in the literature, while in Sec.\,\ref{sec:WGC}, we present the various forms of the WGC. In Sec.\,\ref{sec:rvm}, we review the basic facts of the RVM, its string inspired formulation and vacuumon representation. The swampland criteria in connection to the vacuumon scalar field potential are discussed in Sec.\,\ref{sec:stringSC}.  Our final considerations and conclusions are presented in Sec.\,\ref{sec:conclusions}, where we also demonstrate that the
vacuumon potential satisfies the scalar version of the WGC as presented in \cite{ibanezWGC}.}

\section{The Swampland Conjectures}\label{sec:Swampland}

Motivated by {\it string theory} landscape considerations~\cite{land}, several conjectural criteria, called `the Swampland Conjectures'' (SC) have been proposed in an attempt to consider the embedding of low-energy effective gravitational field theory models, including cosmological ones, that admit de Sitter space-times among their solutions, into consistent microscopic, Ultra-Violet  (UV) complete string/brane theories.  Whether these criteria can be extended beyond string theories is not known, as is also not known whether they truly characterise microscopic string theories themselves, although many concrete stringy examples have been provided that they satrify them.

These criteria can be currently classified into four types, and are all associated with quintessence-type scalar field modes that may give rise to a dS solution of Einstein's equations  (or higher-curvature modifications thereof), that characterise the gravitational sector of the (string-inspired) effective low-energy field theory model:

 \begin{itemize}

\item{(1) \underline{The Field-theory-space Distance Conjecture} (DfsC)~\cite{dSC}}

If an effective low-energy field theory (EFT) of some string theory contains a scalar field $\phi$,  which changes  by a `distance' (in  field space) $\Delta \phi$, from some initial value, then for the EFT to be valid one must have
\be\label{dSC1}
\kappa \, |\Delta \phi | \lesssim c_1~,
\ee
where $c_1 $ is a positive constant of order ${\mathcal O}(1)$, and $\kappa=\frac{1}{\MPl}$ is the gravitational constant in four space-time dimensions, with $\MPl = 2.4 \times 10^{18}$~GeV  the reduced Planck mass.
Violation of the conjecture \eqref{dSC1} implies the `contamination' of the EFT by towers of string states,
not necessarily point-like, which become (ultra) light, as $|\Delta \phi|$ increases with time. Such a descent of ``formerly massive'' string states from the UV would
jeopardise physical conclusions based only on the study of the local EFT .

\item{(2) \underline{The First Swampland Conjecture (FSC)}~\cite{SC1}}

This conjecture provides restrictions on the scalar field(s) self interactions within the framework of a EFT stemming from some microscopic string theory. In particular, for the EFT to be valid,  the gradient  of the scalar potential
$V$ in field space must satisfy:
\be\label{SC1a}
 \frac{|\nabla V|}{V} \gtrsim c_2 \, \kappa > 0
\ee
where $c_2$ is a dimensionless (positive) constant of $\mathcal O(1)$.  The gradient in field space refers to the multicomponent
space of scalar fields $\phi_i$, $i=1,\dots N$ the EFT contains.

The FSC \eqref{SC1a} rules out slow-roll inflation (implying that the standard  slow-roll  parameter $\epsilon = (\sqrt{2}\, \kappa)^{-1} \,  (|V^\prime|/V)$ is of order one),  and with it several phenomenologically successful inflationary models.

It was argued, though, in \cite{SC2a,SC2b} that the FSC may be relaxed, in the sense that consistency of EFT requires
{\it either} the constraint \eqref{SC1a}, {\it or} :

\item{(3) \underline{The Second (Weaker) Swampland Conjecture (SSC)}~\cite{SC2a,SC2b}}

This conjecture states that, if the scalar potential possesses a {\it local maximum}, then {\it near }that local maximum the minimum eigevalue of the theory-space Hessian  ${\rm min}(\nabla_i \, \nabla_j V)$, with  $\nabla_i$ denoting gradient of the potential $V(\phi_j)$ with respect to the (scalar) field $\phi_i$, should satisfy the following constraint:
\be\label{SC2}
\frac{{\rm min}(\nabla_i \, \nabla_j V)}{V} \, \le \, -  c_3 \, \kappa^2 \, < 0
\ee
where $c_3 > 0$ is an appropriate, order $\mathcal O (1)$, positive (dimensionless) constant. The
condition \eqref{SC2} would
be incompatible with the smallness in magnitude of the second of the slow roll parameters $\eta$, as required  for conventional slow-roll single-field inflationary models.

We should stress at this point, that in models for which the SSC applies but {\it not} the FSC, the entropy-bound-based derivation
of FSC~\cite{SC2a} still holds, but for a range of field values outside the regime for which the local maximum of the potential occurs. The {\it critical value/range} on the magnitude of the fields for the entropy-bound implementation of  FSC depends on the details of the underlying microscopic model, as do the parameters $\gamma, b$ appearing in \eqref{eSC1a}.

\item{(4) \underline{Warm Inflation modified First Swampland Conjecture (WImFSC)}~\cite{branden}}

A slight modification of the SSC has been discussed in \cite{branden}, for the case of warm inflation, that is inflationary models for which there is an interaction of the inflaton field $\phi$ with, say, the radiation fields, of energy density $\rho_r$, via:
\bea
&\dot \rho_r + 4 H \, \rho_r = \mathcal Y \, \dot \phi^2 \nonumber \\
& \ddot \phi + (3 H + \mathcal Y) \, \dot \phi + V^\prime = 0,
\eea
where $\mathcal Y$ some (small) positive constant.
The analysis in \cite{branden} shows that the Warm Inflation paradigm is characterised by a slightly modified
FSC bound
\be\label{FSCWI}
\frac{|V^\prime |}{V} \, > \, \frac{2\, \gamma\, b}{3-b} \, \left |1 - \frac{1}{3-\delta} \, \frac{1 - \frac{\kappa^2 \, \dot s}{6\pi \dot H}}
{1  + \frac{\kappa^2 \, s}{6\pi H}} \right |^{-1} \, \kappa,
\ee
where $\gamma, b , \delta > 0$,  with $0 < \delta \le 2$, are constants, whose meaning will become  clear later on ({\it cf.} \eqref{tower}, \eqref{entropy}, below), and $s$ denotes the entropy density of the Universe, which satisfies $ sT \simeq  \frac{\mathcal Y}{3 H}\, \dot \phi^2$,
where $T$ is the temperature. As shown in \cite{branden}, for the explicit models discussed, the deviations
of the bound \eqref{FSCWI} from the one of FSC \eqref{SC1a} are small.

\end{itemize}

The aforementioned Swampland Conditions FSC, SSC or WImFSC,  are all {\it incompatible} with slow-roll inflation, in the sense that in string theory de Sitter vacua seem to be excluded.

\begin{itemize}

\item{(5) \underline{Non-Critical String Modification of the FSC }\cite{emnswamp}}

At this point we remark that in {\it non-critical} (supercritical) string cosmology models, where the target time is identified with the (world-sheet zero mode of the) (time-like) Liouville mode~\cite{emn}, the swampland criteria are severely relaxed, as a result of drastic modifications in the relevant Friedmann equation~\cite{emnswamp}, arising from the non-criticality of the string.
\be\label{swamp}
\frac{|\frac{d}{d \Phi} V (\Phi, \dots ) |}{V (\Phi, \dots)} \, > \, \mathcal O \Big(e^{-|\rm constant | \, \Phi}\Big) \, > \, 0\,,
\ee
where $\Phi$ is the (positive) canonically normalised dilaton of supercritical string cosmologies~\cite{emn},
which increases with cosmic time, and $V(\Phi, \dots) > 0$ its potential, which is non trivial for non-critical strings~\cite{aben,emn}. Thus the constraint \eqref{swamp} is trivially satisfied for long times after an initial time, for which $\Phi$ is large.
For such (non-equilibrium, relaxation) cosmologies, slow-roll inflation is still valid.

\end{itemize}

In the present article we shall only be concerned with cosmological models that are embeddable in critical string theory, and we shall examine the validity of  the SC, as formulated above, in the context of \joantext{the aforementioned running vacuum effective cosmological field theory, or Running Vacuum Model (RVM) for short, see\,\cite{JSPRev2013,JSPRev2015} for a review.  The main traits of the RVM will  briefly described  in section \ref{sec:rvm}.}

Before embarking on such an analysis, we stress that the above SC are not entirely independent of each other. In particular, the constraint \eqref{SC1a} can also be derived by the DfsC, by employing
entropy considerations in de Sitter  space time~\cite{SC2a}. Indeed, let one
consider  a (quintessence type) homogeneous scalar field $\phi(t)$, assumed for concreteness to be an increasing function of (cosmic) time $t$. Our convention is that the derivative of the potential with respect to the field $\phi$ is {\it negative}.
As the distance of the field $\Delta \phi$ from an initial value grows with time, then, according to the DfsC, a tower of massive string states $N(\phi)$ acquires light masses $m \sim \exp (-a |\Delta \Phi|)$, where $a > 0$ is a constant (with mass dimension $-1$), which depends on the details of the theory.  The effective light degrees of freedom $N(\phi)$ is a function of the field $\phi$. At any moment in the cosmic time $t$, they are parametrised as~\cite{SC1}:
\be\label{tower}
N(\phi) = n(\phi)\, \exp (b \, \kappa \, \phi)
\ee
where $b$, like $a$, is another positive constant that depends on the mass gap and other details of the underlying string theory. Given that, according to the DfsC, the number of string states that become light increases with increasing $\phi$, we conclude that the (positive) function $n(\phi) > 0$ in \eqref{tower} must be a monotonically increasing function of $\phi$, {\it i.e}. $\frac{d n(\phi)}{d \phi} > 0$.

In an accelerating  almost de Sitter Universe,  with an approximately constant Hubble parameter $H$, of interest to us here, the entropy of this tower of string states is an increasing function of the Hubble horizon $1/H$:
\be\label{entropy}
S_{\rm string~states} (N, H^{-1}) = N^\gamma \, (\kappa H)^{-\delta}
\ee
where $\gamma, \delta > 0$. From the studied examples in string theory, the authors of \cite{SC2a} argued that $0 < \delta \le 2$.  The presence of the Hubble Horizon of area $A=4\pi H^{-2}$ for this expanding Universe, implies,
according to Bousso's covariant entropy bound applied to a cosmological background~\cite{Bousso}, that
the entropy \eqref{entropy} is bounded from above as:
 \be\label{bousobound}
 S_{\rm string~states} (N, H^{-1}) \le \nicktext{2\pi A M_{\rm Pl}^2=2\pi A\kappa^{-2}}\,.
 \ee
The right-hand side of the above inequality is the  \joantext{Bekenstein}-Gibbons-Hawking entropy~\cite{Bekenstein,Hawking,GH}, so, on account of \eqref{entropy}, one obtains:
\be\label{GH}
N^\gamma \, (\kappa H)^{-\delta}  \le 8\pi^2 (\kappa H)^{-2}.
\ee
If the potential energy of the scalar field dominates the energy density of the Universe, {\it i.e.} the kinetic energy
of the scalar field responsible for inflation is ignored in  front of the potential energy, then, upon using  the Friedmann equation, one may rewrite \eqref{GH} in the form
\be
\frac{\kappa^4 \, V}{3} = \Big(\frac{8\pi^2}{N^\gamma}\Big)^{2/(2-\delta)}.
\ee
Taking into account that the derivative of the potential with respect to the field $\phi$ is negative, we obtain, after some straightforward steps,
\be\label{eSC1b}
\frac{|V^\prime |}{V} \ge  \frac{2}{2-\delta} (\ln [N^\gamma])^\prime,
\ee
where $V^\prime (\phi) = \frac{dV(\phi)}{d\phi}$. On account of \eqref{tower}, then, one obtains
\be\label{eSC1a}
\frac{|V^\prime|}{V} \, \ge \, \frac{2\, \gamma}{2-\delta} \, \Big(\frac{n^\prime}{n}
 + b \, \kappa \Big) \, > \, \frac{2\, b\, \gamma}{2-\delta} \, \kappa \, ,
\ee
where, to arrive at the last inequality, we took into account that $n^\prime > 0$. Upon comparing \eqref{eSC1b} with \eqref{SC1a}, one obtains the
FSC with $c_2 = \frac{2\, b\, \gamma}{2-\delta} > 0$, provided that the parameters $b\, ,\gamma $ and $\delta$ are such that  $c_2$ is of order $\mathcal O(1)$.  If only point-like states are included in the tower of states $N(\phi)$, then $\delta =0$.

\section{The Weak Gravity Conjectures}\label{sec:WGC}

We conclude the \nicktext{introductory part of our work} by mentioning that another conjecture that was put forward~\cite{WGC}
as a requirement for the consistent embedding of low-energy string effective theories from the string `landscape' into microscopic quantum gravity was that of the so-called `\emph{Weak Gravity Conjecture}' (WGC). \nicktext{WGC essentially states} that a consistent EFT, which might contain extra forces,  should respect the fact that gravity is the `weakest' of all forces. We note that there are various forms of this conjecture, which is still plagued by many ambiguities and specific dependence on physical processes involved, involving scalar fields~\cite{Palti,ibanezWGC,kusenko,benakli,ibanez2}, but the advantage is that the WGC might go beyond string theory and dS space-times, and be seen as a generic, but yet incomplete, attempt towards the formulation of model independent criteria for embedding EFT into quantum gravity frameworks. The ambiguities in its present formulation reflect of course our current ignorance on the elusive microscopic theory of quantum gravity.

For our purposes in this work, the form of WGC for scalar fields we shall use is the one in \cite{ibanezWGC} and its generalization in \cite{kusenko}.
This form of WGC for scalars expresses essentially the requirement that, if $V(\phi)$ is the potential of some scalar field $\phi$, then the `force' that it corresponds to should be {\it no weaker} than gravity. By considering appropriate four-point scattering amplitudes of this scalar, including both tree-level graphs mediated by the scalar itself, as well as (repulsive)
contact interactions (that include massive mediators that may exist in the EFT), and comparing them with the corresponding amplitudes mediated by gravity, the authors of \cite{ibanezWGC} have conjectured the following relation for the scalar WGC to be maintained:
\begin{align}\label{WGC}
\kappa^2 (V^{\prime\prime})^2 \, \le \, 2\, (V^{\prime\prime\prime})^2 - V^{\prime\prime} \, V^{\prime\prime\prime\prime} \,,
\end{align}
where the left-hand side of this inequality represents the gravity-mediated graph, whilst on the right-hand-side, the first term corresponds to the scalar mediated amplitudes and the last term to the contact scalar interactions, that effectively include massive mediators. Unlike other forms of the scalar WGC~\cite{Palti}, the conjecture \eqref{WGC} allows for axion potentials to be included in a self consistent way.

We stress that there is no diagrammatic proof of this relation, and the coefficients of the various terms of the four-point scalar amplitudes used, were the minimal  ones. Indeed, we mention that
the relation was generalised in \cite{kusenko} to include positive coefficients of $\mathcal O (1)$
in front of the terms of the right-hand side of the inequality \eqref{WGC}, and in this way Q-ball and other field-theory solitonic-configuration potentials could be made compatible with the WGC.

In our work below \nicktext{(see section \ref{sec:conclusions})}  we shall demonstrate that the RVM is indeed compatible with \eqref{WGC}, along with the SC, by using a scalar field representation of some aspects of RVM, by means of the so-called  ``vacuumon' field'~\cite{vacuumon,rvmsugra}.
We now proceed to review briefly this model by paying attention to its important features that we shall make use of in this work.

\section{The Running Vacuum Model (RVM):  a brief review of its most important features} \label{sec:rvm}

One of the most interesting alternatives to the standard concordance $\Lambda$CDM models is the
``running vacuum model'' (RVM) of the Universe,  \joantext{see \cite{JSPRev2013,JSPRev2015} for a review. Such a running means that the model can provide the connection between the  values of the vacuum energy density from one energy scale to another throughout the cosmological history. \nicktext{This feature was actually} originally suggested long ago from the point of view of the renormalization group in curved spacetime from different perspectives\,\cite{ShapSol1,Fossil2008,ShapSol2}. Interestingly enough, it can provide also a framework for the possible time variation of the so-called fundamental constants of nature\,\cite{FritzschSola1,FritzschSola2}.
Such a model  has also been argued to provide a uniform effective description of the Universe evolution from inflation till the present era~\cite{rvmhistory,rvmThermal}; see also \cite{solaentropy} and the long list of references therein. As previously mentioned, the model provides a competent
fit  to the plethora of the cosmological data and contributes to the alleviation of recent tensions with predictions of the $\Lambda$CDM model ~\cite{rvmpheno}.}

\subsection{RVM Cosmic Evolution \label{sec:evol}}

\joantext{The RVM  evolution of the Universe is based on the following general form of the vacuum energy density  in terms of powers of the Hubble parameter $H$ and its first cosmic-time derivative $\dot H$~\cite{JSPRev2013,JSPRev2015}:}
\be
\rvm(H,\dot{H})
=a_0+a_1\,\dot{H}+a_2\,H^2+a_3\,\dot{H}^2+a_4\,H^4+a_5\,\dot{H}\,H^2+ \joantext{a_6H\ddot{H}}\dots,
\label{GRVE}
\ee
\nicktext{where the overdot denotes cosmic-time derivative,} \joantext{the (real) coefficients $a_i$ have different dimensionalities in
natural units, and  the $\dots$ denote the  possible decoupling terms (suppressed by mass powers), which} \nicktext{are irrelevant for our discussion}.  \joantext{The structure of some of these terms was hinted long ago  in \,\cite{ShapSol1}  from the point of view of the renormalization group (RG), and were further elaborated in\,\cite{Fossil2008,ShapSol2} --  see also \cite{ShapSolStefan2005,Babic2005,Maggiore2011,FritzschSola1} for different interpretations and phenomenological applications.  A first connection with an action formulation was actually provided in \,\cite{Fossil2008} in the context of anomalous conformal field theories.  Very recently, the computation of the vacuum energy density of a non-minimally coupled scalar field in the spatially flat FLRW background has been presented in the comprehensive work\, \cite{Cristian2020}. In it,  the low energy terms proportional to $H^2$ and $\dot{H}$  in Eq.\,\eqref{GRVE} as well as most of the higher order terms ${\cal O}(H^4)$ have been explicitly derived for the first time.  The higher order terms  which appear in that calculation adopt one of the three forms $\dot{H}^2,\, H^2 \dot{H}$  and  $ H\ddot{H}$,  all of them of adiabatic order $4$, which stem from varying  the higher order covariant terms  ${\cal O}(R^2)$ existing in the vacuum action and expressing the renormalized result in the FLRW metric.  The method employed in \, \cite{Cristian2020} is a variant of the adiabatic regularization and renormalization of the energy-momentum tensor\,\cite{ParkerToms}. Notice that only the even adiabatic orders appear in the final result obtained in that calculation, which explains why only the terms involving an even number of time derivatives of the scale factor appear in Eq.\,\eqref{GRVE}. This is, of course,  a necessary condition for the general covariance of the result,  see  \cite{Cristian2020} for technical details and related references.  As mentioned, all the terms quoted in \eqref{GRVE}  appear in that  QFT calculation, with one single relevant exception, which is the term proportional to $H^4$  (with no derivatives). As it turns} \nicktext{out}, \joantext{this kind of terms seem to be more characteristic of the string-inspired RVM formulation, where they can be generated through the gravitational anomalous condensates triggering the inflationary epoch --- see \cite{bms1,bms2,bms3}.  The $\sim H^4$-terms are precisely the ones on which we will focus our attention hereafter.}

For the evolution of the Universe, from the very early (inflationary) stages to the current era,
a simplified structure of the  general RVM expression \eqref{GRVE} suffices, based on the approximation that
at various epochs, with a constant decceleration parameter $q$ per era, one can write
\be\label{Hdot}
\dot H \simeq - (q + 1) \, H^2
\ee
and thus in practice one can use ~\cite{rvmhistory,solaentropy}:
\begin{equation}\label{rLRVM}
\rho^{\Lambda}_{\rm RVM}(H) = \frac{\Lambda(H)}{\kappa^2}=
\frac{3}{\kappa^2}\left(c_0 + \nu H^{2} + \alpha
\frac{H^{4}}{H_{I}^{2}} + \dots \right) \;,
\end{equation}
where $\dots$ denote terms of order $H^6$ and higher, and we used the notation $\Lambda (H)$ to stress the connection of the
RVM with a ``\joantext{running cosmological term}'' with an equation of state \joantext{(EoS)}  identical to that of a cosmological constant:
\be\label{wl}
w_{\rm RVM} = -1.
\ee
\joantext{As previously noted},  the dependence of $\rho^\Lambda_{\rm RVM}(H)$
on even powers of $H$ is the result of general covariance~\cite{JSPRev2013}.
In \eqref{rLRVM},
 $H_I$ denotes the Hubble parameter around the  GUT scale, and
$c_0$ is an integration constant (with mass dimension $+2$ in
natural units), \joantext{which is numerical close (for $|\nu|\ll1$) to the  present day value of the cosmological constant.}

One may consider matter/radiation excitations of the running vacuum, which are described by the following
cosmological (Friedmann) equations in the presence of a running $\Lambda (t)$:
\be
 \kappa^{2}\rho_{\rm tot}=\kappa^2 \rho_{m}+\Lambda (t)= 3H^2 \;,
\label{friedr} \ee
\be
\kappa^{2}p_{\rm tot}=\kappa^2 p_{m}-\Lambda (t) =-2{\dot H}-3H^2,
\label{friedr2}
\ee
where the overdot denotes derivative with respect to cosmic time
$t$, and
$\rho_{\rm
tot}=\rho_{m}+\rho_{\Lambda}$ and $p_{\rm
tot}=p_{m}+p_{\Lambda}=p_{m}-\rho_{\Lambda}$, are the total energy density and pressure density
of both the vacuum (suffix $\Lambda$) and matter/radiation (suffix $m$) terms.
Specifically, the quantity
$\rho_{m}$ denotes the density of matter-radiation, while
$p_{m}=\omega_{m} \rho_{m}$ is the corresponding pressure, and
$\omega_m$ is the EoS for matter/radiation components ($w_{m}=0$ for nonrelativistic matter and
$w_{m}=1/3$ for  radiation). On the other hand, as already mentioned, the EoS of the RVM is \eqref{wl}, i.e. $\rho_\Lambda = - p_\Lambda$.

It is important to note that, unlike the standard  $\Lambda$CDM model of cosmology,  where $\Lambda=$const.,
in the RVM \joantext{there are} non trivial interactions between radiation/matter and
vacuum, which are manifested in the modified conservation equation for the matter/radiation energy density $\rho_m$, obtained from the corresponding Bianchi identities of the RVM Universe:
\begin{equation}
\dot{\rho}_{m}+3(1+\omega_m )H\rho_{m}=-{\dot{\rho}}^{\Lambda}_{\rm RVM}\,,
\label{frie33}
\end{equation}
with $\omega_m$ the EoS of the matter/radiation fluid, to be distinguished from that of the vacuum \eqref{wl}.

We note at this stage that, in view of \eqref{Hdot}, the modifications in the right-hand side of the conservation equation \eqref{frie33} are of order $H^3$ and higher, and also suppressed by factors $q + 1$ which, during the inflationary phase we shall be interested in, are almost zero (in fact, during the inflationary phase, for which the $H^4$ term in \eqref{rLRVM} dominates, one has
$\dot \rho^\Lambda_{\rm RVM} =  \mathcal O \Big((q+1) H^5\Big)$). Thus, such modifications are strongly suppressed in our case, and, like the warm inflation scenario~\cite{branden} ({\it cf.} \eqref{FSCWI}), they will {\it not} affect the formulation of the SC.

Taking into account the RVM expression \eqref{rLRVM}, and  using the equations \eqref{friedr2} one can obtain  from
\eqref{frie33} a solution for $H(a)$ as a function of the scale factor and the equations of state of `matter' in RVM~\cite{rvmhistory}:
\begin{equation}\label{HS1}
 H(a)=\left(\frac{1-\nu}{\alpha}\right)^{1/2}\,\frac{H_{I}}{\sqrt{D\,a^{3(1-\nu)(1+\omega_m)}+1}}\,,
\end{equation}
where $D>0$ is the integration constant. \joantext{On assuming $|\nu|\ll1$}, we then observe
that for early epochs of the Universe, where the scale factor $a \ll 1$, one has
$D\,a^{3(1-\nu)(1+\omega_m)} \ll 1$, and thus
an (unstable) De Sitter phase~\cite{rvmhistory} characterised by
an approximately constant $H \simeq \left(\frac{1-\nu}{\alpha}\right)^{1/2}\,\, H_{I}$.

A smooth evolution from the early to late stages of the Universe is then assumed in generic RVM approaches,
\joantext{with the coefficients $ 0<|\nu, \alpha|\ll1$ in \eqref{rLRVM} staying {\it the same} in all eras}.
As already mentioned, in the early Universe, the term $H^4$ dominates in \eqref{rLRVM}, \joantext{and drives inflation and all basic thermodynamical aspects of the early uniuverse~\cite{rvmhistory,solaentropy}}, without the need for a fundamental  inflaton field, while at late eras the terms
$c_0$ and $H^2$ dominate, and make a prediction on the deviation of the current dark energy from that of the standard concordance $\Lambda$CDM model. Standard late-epoch CMB and other current cosmological data phenomenology imply $\nu \sim 10^{-3}$\cite{rvmpheno}, a result which is consistent with previously existing theoretical estimates\,\cite{Fossil2008}, while inflationary considerations lead to the rather generic conclusion that $H_I$ should be near the GUT scale, \joantext{with $|\alpha|<1$.}
In generic RVM models, one makes the assumption that the matter ($m$) content of the theory at early epochs consists of relativistic particles \joantext{with an equation of state $w_m =1/3$. In such a case, for $|\nu| \ll 1$} one obtains from
\eqref{HS1}:
\begin{equation}\label{HS1rad}
 H(a)=\left(\frac{1}{\alpha}\right)^{1/2}\,\frac{H_{I}}{\sqrt{D\,a^{4}+1}}\,,
\end{equation}
and  for the early (unstable) de Sitter phase one has
$Da^{4} \ll 1$, and $H$ remains approximately constant.  \nt{This expression is appropriate for the early universe, i.e. for  $a\simeq 0$.  A better understanding of its meaning in the early universe and a more comfortable connection with the current universe ($a\simeq 1$) can be made more manifest by rescaling some quantities in it. In particular, it is convenient to eliminate $D$ in terms of a more physical parameter of the very early universe (essentially the point where inflation stops). To determine this point we need to compute the matter and vacuum energy densities as well.  The  equality point  $a_{\rm eq}$  of (relativistic) matter and vacuum energy densities (defined by the condition $\rho_r(a_{\rm eq})=\rho_\Lambda(a_{\rm eq})$) can then be used to eliminate $D$.  We define also for convenience the rescaled variable
 \begin{equation}\label{eq:ahat}
  \ha\equiv \frac{a}{\astar}\,,
\end{equation}
where $\astar$ is related to $a_{\rm eq}$ through\,\cite{solaentropy}
\begin{eqnarray}\label{aeq0}
D=\frac{1}{1-2\nu}\,a_{\rm eq}^{-4(1-\nu)}\equiv\astar^{-4(1-\nu)}\,.
\end{eqnarray}
Quite obviously,   $\astar$ is essentially equal to $a_{\rm eq}$ since $|\nu|\ll1$ but it is more convenient to use the former since the formulas simplify.
With these definitions, the more appropriate form for $H$ and the associated energy densities of matter and vacuum energy read as follows\,\cite{rvmhistory,solaentropy}:}
\begin{eqnarray}\label{hubbleeq0}
H(\ha)=\frac{\tHI}{\sqrt{1 +\ha^{4(1-\nu)}}}\,,
\end{eqnarray}
\begin{eqnarray}\label{eq:rhorfinal}
\rho_r(\ha)&=&\trI (1-\nu)\,\frac{\ha^{4(1-\nu)}}{\left[1+  \ha^{4(1-\nu)}\right]^2}
\end{eqnarray}
and
\begin{eqnarray}\label{eq:rhoLfinal}
\rho_\Lambda(\ha)&=&\trI\, \frac{1+\nu \ha^{4(1-\nu)}}{\left[1+  \ha^{4(1-\nu)}\right]^2}\,.
\end{eqnarray}
In the above equations we have also defined a rescaled form for $H_I$ and $\rho_I$ :
\begin{equation}\label{eq:tHI}
\tHI=\sqrt\frac{1-\nu}{\alpha}\,H_I\,,\ \ \ \ \ \ \trI=\frac{3}{\kappa^2}\,\tHI^2\,.
\end{equation}
\nt{As we can see from  (\ref{eq:rhoLfinal}), the value of $\trI$ is nothing but the vacuum energy density at $a=0$:  $\rL(0)=\trI$, hence at the beginning of the inflationary epoch.  }

\nt{The above equations clearly show the transfer of energy from  vacuum decay to matter.  At $\ha=0$, the vacuuem energy is maximal whereas the matter density is zero. From this point onwards the process continues until reaching a balance at $a_{\rm eq}$, where $\rho_r(a_{\rm eq})=\rho_\Lambda(a_{\rm eq})$.  We can derive the numerical order of magnitude  of the point $a_{\rm eq}$ $\simeq \astar$ by  taking into account that,  in the asymptotic limit ($\ha\gg1$, i.e.  $a\gg a_{\rm eq}$),  hence deep into the radiation epoch, the radiation density (\ref{eq:rhorfinal}) behaves as
\begin{eqnarray}\label{eq:rhorfinal2}
\rho_r(a)&=&\trI (1-\nu) \ha^{-4(1-\nu)}=\trI (1-\nu) \astar^{4(1-\nu)}\,a^{-4(1-\nu)}\,.
\end{eqnarray}
As we can see, we are able to recover from \eqref{eq:rhorfinal} the standard behavior of the radiation density in the asymptotic limit, $\rho_r(a)\sim\rho_{r0} a^{-4(1-\nu)}$, up to a tiny correction in $\nu$.
So both equations must be the same and both must reproduce the same radiation density at present: $\rho(a=1)=\rho_{r0}$.  This provides a normalization point to fix a parameter.  On the other hand the energy density at the inflationary period must be of order of the GUT one, $\trI\sim M_X^4$, with $M_X\sim 10^{16}$ GeV the typical GUT scale. Using this fact  and the current value of the  radiation energy density in units of the critical density, $\Oro=\rro/\rco\sim 10^{-4}$, it is easy to derive $a_{\rm eq}\simeq\astar\sim  10^{-29}$\,\cite{solaentropy}, which indeed places the equality point between radiation and vacuum energy in the very early universe, essentially at the end of inflation\,\footnote{\nt{Compare  with  the equality point between radiation and nonrelativistic matter:  $a_{\rm EQ}\sim 3\times 10^{-4}$. }}}.

\nt{However, as we discussed in \cite{bms1,bms2}, and shall  further address below, in the context of a specific string-inspired RVM model the `matter content' is different from that of relativistic matter, and moreover there is no such perfectly smooth evolution from the de Sitter inflationary eras to the current era, as there are phase transitions at the exit from inflation, which result in new degrees of freedom entering the EFT, although qualitatively the main features of RVM are largely preserved.  In fact, the above discussion shows that a smooth evolution can lead to a  reasonable picture, in which the the standard radiation dominated epoch ($\rr\sim a^{-4}$) follows continuously from the inflationary one. A more realistic scenario, however, requires an intermediate step (phase transition)  in which the  Kalb-Ramond (KR)  axion from the effective low-energy string theory (see next section) will play a significant role, see Sec.\,\ref{sec:string}. Needless to say, this is an important point of the stringy version of the RVM (which was absent in its original form) and that  is under discussion here.}

\nt{The following observation may be in order here so as to  better clarify the connection between the RVM physics of the early universe with the one expected at present.  From the generic RVM expression for the vacuum energy density \eqref{rLRVM}, one might expect that the connection with the current universe is obtained in the limit  $\alpha\to 0$. However, such limit is undefined for both the Hubble rate \eqref{HS1rad} as well as  for the energy densities \eqref{eq:rhorfinal}-\eqref{eq:rhoLfinal} and hence it cannot really be implemented ~(see \cite{solaentropy} for details).  Indeed, a crucial virtue of the RVM approach is that the initial value of the Hubble rate, $H(0)=\tHI\simeq {H_I}/\sqrt{\alpha}$ is finite and hence there is no singular initial point. To insure this feature, it is indispensable that $\alpha>0$ (strictly).  In the limit where we reach the vanishing value $\alpha=0$ the entire RVM physics of the early universe disappears since  no non-singular solution  can exist at $a=0$,  except the trivial one  ($H=0$), as can be easily checked.  In other words, it is only when the term $H^4$ is present, and carrying a positive coefficient, that nonsingular solutions to that equation can exist.  A nonvanishing value for $\alpha$ is mandatory and hence the way to connect the early universe and the current universe in the context of the RVM model \eqref{rLRVM} is \textit{not} by performing a zero limit of the parameters $\nu, \alpha$  but by just letting the evolution of $H$ to interpolate between the different epochs.  The  two coefficients must be present and nonvanishing in the entire cosmic history.  The connection between epochs  is implemented dynamically through  the relative strength of  $H^4$ vs $H^2$  that changes when moving from early epochs to the current ones, in which  the former term is completely negligible compared to the latter.  The function \eqref{rLRVM} is indeed a continuous function of $H$ and  moves from $H^4$ dominance into $H^2$ dominance, and finally we are left with a mixture of constant (and dominant) term plus a tiny correction $\sim\nu H^2$. This means that, according to the RVM,  the dark energy in the current universe is evolving, as  there is still a mild dynamical vacuum energy $\sim H^2$  on top of the dominant term (the cosmological constant).  Although it may create the illusion of quintessence,  it is just residual dynamical vacuum energy that helps to improve the fit to the data\,\cite{rvmpheno}.}

\nt{While this is the standard picture  within the RVM\,\cite{JSPRev2013,JSPRev2015},  in a stringy RVM formulation the contributions to the current-era cosmological constant may come from condensation of much-weaker GW, and the evolution cannot be described by a smooth solution \eqref{HS1}, connecting the initial inflation to the current epoch\,\cite{bms1,bms2}. More details will be given here.  Basically, the GW condensation leading to the initial and current-era (approximately) de Sitter space times are viewed as dynamical phase transitions, whose presence affect the smoothness of the evolution of the stringy Universe. In this respect, the RVM can be seen as providing an effective description within each epoch, with non-trivial coefficients of the various $H$-powers in the string-inspired RVM analogue of \eqref{rLRVM}, which are computed microscopically in the various eras, as we discussed in \cite{bms1} and revisit below.}

\nt{After  introducing the basics of the two RVM versions (standard and stringy), we now proceed to discuss these matters, as this is important for our main point of the current study, namely the compatibility of the RVM with swampland.}

\subsection{Embedding RVM in String Theory \label{sec:string}}

In \cite{bms1,bms2,bms3} we have derived the RVM within the context of four-dimensional string-inspired cosmological model, based on critical-string low-energy effective actions of the graviton and antisymmetric tensor (spin-one) Kalb-Ramond (KR) fields of the massless (bosonic) string gravitational multiplet~\cite{gsw,string,kaloper}. Crucial to this derivation was the
KR field, which in four space-time dimensions is equivalent to a pseudoscalar massless excitation, the KR axion field $b(x)$. Such a field couples to gravitational anomalies, through the effective low-energy string-inspired gravitational action, which in \cite{bms1,bms2,bms3} has been assumed to describe fully the early Universe dynamics:
\begin{align}\label{sea4}
S^{\rm eff}_B =\; \int d^{4}x\sqrt{-g}\Big[ -\dfrac{1}{2\kappa^{2}}\, R + \frac{1}{2}\, \partial_\mu b \, \partial^\mu b
+   \sqrt{\frac{2}{3}}\,
\frac{\alpha^\prime}{96 \, \kappa} \, b(x) \, R_{\mu\nu\rho\sigma}\, \widetilde R^{\mu\nu\rho\sigma} + \dots \Big]~,
\end{align}
where the last term in the right-hand side of this equation is
a CP-violating gravitational anomaly \joantext{(or gravitational Chern-Simons term)}.  The latter is affected by the presence of CP-violating primordial
gravitational waves (GW) in the early Universe~\cite{stephon}, which can condense leading to a RVM type
vacuum energy density~\cite{bms1,bms2}:
\begin{align}\label{toten}
\rho_{\rm total} &  \simeq
3\kappa^{-4} \, \Big[\newtext{ -1.65} \times 10^{-3} \Big(\kappa\, H \Big)^2
+ \frac{\sqrt{2}}{3} \, |\overline b(0)| \, \kappa \, \times {5.86\, \times} \, 10^6 \, \left(\kappa\, H \right)^4 \Big] > 0~.
\end{align}
The vacuum energy density $\rho_{\rm total}$
receives contributions from the KR axion  and gravitational wave spin two fluctuations (the dilaton is assumed to have been stabilised to a constant term~\cite{bms1,bms2}, although it imposes constraints due to its equations of motion). \joantext{The vacuum energy density structure \eqref{toten} predicted by such a string-inspired framework happens to be of the RVM type \eqref{rLRVM}}, but here, in contrast to the conventional RVM\cite{JSPRev2013}, the $\nu$ coefficient of the $H^2$ term is {\it negative}, due to negative contributions from the gravitational Chern-Simons anomaly term, which overcome the positive $\nu$ contributions from the `stiff' KR axion -- \joantext{see Eq.\,(\ref{stiff}) below}~\cite{bms1,bms2}. In the early Universe the dominant term is the $H^4$ term, with a positive coefficient, arising from the GW condensate, and thus the energy density $\rho_{\rm total}$ is {\it positive} and drives an almost de Sitter (inflationary) phase in that period. Here $\overline b(0)$ indicates an initial value of the background KR field (a solution of the pertinent equations of motion). Phenomenology requires $|\overline b(0)| \gtrsim 10 \MPl$. \joantext{Since $H_I^2=(\kappa^2/3)\rho_X$,  with $\rho_X\sim M_X^4$, one can easily check that the corresponding coefficient $\alpha$ in \eqref{rLRVM} is of order $0.1$ for a typical GUT scale $M_X\sim 10^{16}$ GeV.}

At the inflationary exit period, massless chiral fermionic matter, as well as gauge degrees of freedom, are assumed to be created~\cite{bms1,bms2}, which enter the effective action via the appropriate fermion kinetic terms and interaction with the gravitational and gauge anomalies.
\joantext{The primordial gravitational anomaly terms} \nicktext{ {\it  are cancelled} by the chiral matter contributions~\cite{bms1,bms2}}, \joantext{but the triangular (chiral)  anomalies (electromagnetic and of QCD type) in general remain. This must be so, of course,  since some physical processes directly depend on them.}

In the post inflationary phase
the KR axion acquires, through instanton effects, a non perturbative mass, and may play the role of Dark Matter.
It can be shown that the late-era vacuum energy density acquires a standard RVM form \eqref{rLRVM}, with positive
coefficient $\nu_{\rm late} \sim \mathcal O(10^{-3})$,  consistent with phenomenology. At late eras, higher than $H^2$ terms in the energy density are not phenomenologically relevant and thus can be safely ignored. The $\nu_{\rm late} H^2$ corrections to the standard current-era cosmological constant term $c_0$ lead to distinctive signatures of a ``running'' dark energy, which helps to alleviate the aforementioned tensions in the cosmological data with the predictions of the
standard $\Lambda$CDM\,\cite{rvmpheno,snowmass2020}.

Thus, in contrast to the standard RVM, the string-inspired RVM cosmological evolution appears to be {\it not}
 smooth, due to the involved {\it phase transitions} at the exit from the inflationary phase, which result in extra degrees of freedom entering the effective theory. But this does not affect the basic features of the RVM, of inducing inflation at the early eras, through its $H^4$ term, without the need for an external inflaton degree of freedom, and a running dark energy at the current era, through the $H^2$ terms.

An important novel feature of the  string-inspired RVM model~\cite{bms1,bms2}
concerns its `matter' content, which  comes from the massless gravitational axion $b(x)$ field, which has a `stiff matter' EoS~\cite{stiff,stiff2}
\be\label{stiff}
w^{\rm stiff}_{\rm m-string-RVM}=+1\,,
\ee
 in contrast to the assumption of relativistic matter ($w_m=+1/3$) made in generic phenomenological RVM models of the early universe~\cite{rvmhistory} and \cite{vacuumon}, as discussed above. Thus, during the early \joantext{de Sitter} era, in our case, and in view of the fact that \joantext{$|\nu|={\cal O}(10^{-3}) \ll 1$} in \eqref{toten}, one would have from \eqref{HS1}
\be\label{stringHdS}
H(a) _{\rm early~string~RVM} \simeq
\left(\frac{1}{\alpha}\right)^{1/2}\,\frac{H_{I}}{\sqrt{D_{\rm string} \,a^{6}+1}}\,,
\ee
to be compared with \eqref{HS1rad}.

 \newtext{Thus, in this scenario, the ordinary radiation-dominated phase of the Universe follows} \joantext{ the first period of vacuum decay into massless axions.} \nicktext{During the radiation phase,} \joantext{ the axions acquire a mass  as a result of non-perturbative instanton effects, and thus can play a r\^ole as  DM candidates in the current universe\,\cite{bms1,bms2}.} \newtext{It should be remarked that, in this respect, our considerations are in qualitative agreement with the ideas of \cite{stiff}, \joantext{except that here we have a stiff fluid made of axions whereas in that work it is a phenomenological cold gas of baryons}.}

\subsection{The ``Vacuumon'' Representation of the Running Vacuum Model : early Eras \label{sec:vacuumon}}

On combining Friedmann's Eqs.(\ref{friedr})-(\ref{friedr2}) and
following standard approaches~\cite{rvmpheno}, using a classical scalar field representation of the
total energy/pressure densities of the RVM framework in the presence of matter/radiation~\cite{vacuumon} (see also \cite{rvmsugra} for an earlier discussion):
\be\label{sfrvm}
\rho_{\rm tot}\equiv \rho_{\phi}={\dot \phi}^{2}/2+U(\phi), \quad
p_{\rm tot}\equiv p_{\phi}={\dot \phi}^{2}/2-U(\phi).
\ee
we obtain
\begin{align}
\dot{\phi}^{2} =-\frac{2}{\kappa^{2}}\dot{H} \quad \Rightarrow \quad
\phi= \pm \frac{\sqrt{2}}{\kappa}\int
\left(-\frac{H^{'}}{aH}\right)^{1/2}da\;,
\label{ff3}
\end{align}
and
\be\label{potH}
U =\frac{3H^{2}}{\kappa^{2}}\left(1+\frac{a}{6H^2}\,\frac{d H^2}{da}\right) \;.
\ee
In terms of this field $\phi$, termed the ``vacuumon''~\cite{rvmsugra,vacuumon},
and using \eqref{HS1rad} or \eqref{stringHdS}, depending on the RVM matter content at early eras, and \eqref{potH},
one can then describe the early $H^4$-dominated early Vacuum phase by means of an effective  potential $U(\phi)$.

We shall consider below the two cases of RVM separately, for the purposes of comparison:
\begin{itemize}
\item{\bf (i)} the generic RVM,
assuming relativistic matter,
\begin{align}
\label{Pott}
U(\phi) &=U_{0}\; \frac{2+{\rm
cosh}^{2}(\kappa \phi)} {{\rm cosh}^{4}(\kappa \phi)}  \;,  \\
U_{0} &=\frac{H^{2}_{I}}{\alpha \kappa^{2}}\;, \\
\kappa \, \phi (a) &= {\rm sinh}^{-1} (\sqrt{D} a^2) = \ln \Big(\sqrt{D}\, a^2 + \sqrt{D\, a^4 + 1}\Big),
\nonumber
\end{align}

\item{\bf (ii)} the specific string-inspired RVM~\cite{bms1,bms2}, with `stiff' gravitational axion $b$ `matter' at early epochs, \eqref{stiff}.
\begin{align}
\label{Pott_string}
U(\phi) &= \widetilde U_{0}\; \frac{\tfrac{2}{3}+{\rm
cosh}^{2}(\kappa \phi)} {{\rm cosh}^{4}(\kappa \phi)}  \;,\\
\widetilde U_{0}&  =\frac{9H^{2}_{I}}{\alpha \kappa^{2}}\;, \\
\kappa\, \phi (a) &= \sqrt{\frac{2}{3}}\, {\rm sinh}^{-1} (\sqrt{D_{\rm string}} a^3) = \sqrt{\frac{2}{3}}\,\ln \Big(\sqrt{D_{\rm string}}\, a^3 + \sqrt{D_{\rm string} \, a^6 + 1}\Big).
 \nonumber
\end{align}

\end{itemize}

\begin{figure}
\includegraphics[width=0.45\textwidth]{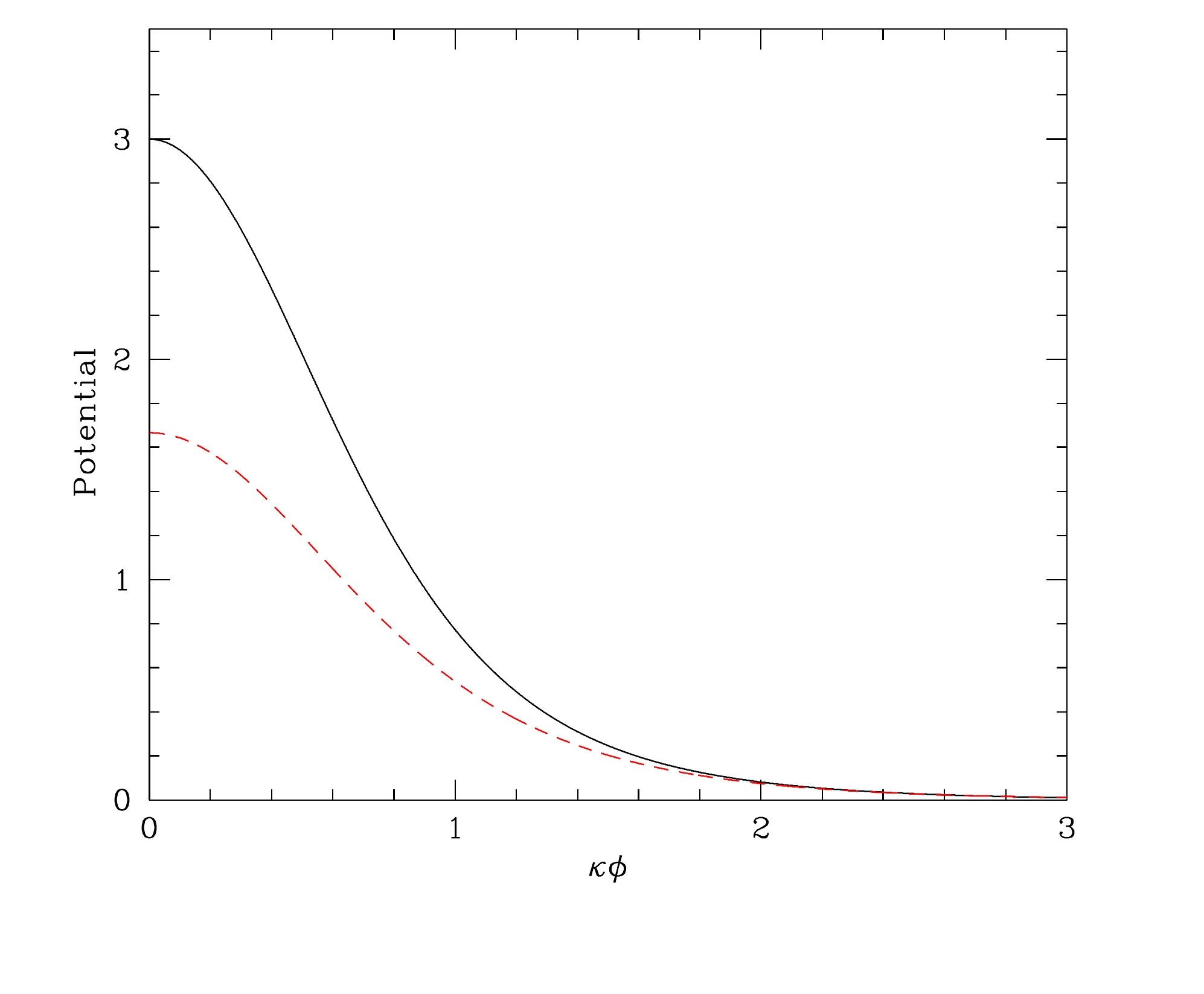}
\caption{{\it The ``vacuumon'' potential $U/U_{0}$ for a classical scalar field representation of the early-Universe RVM.
{\underline \it Solid line} : Generic RVM~\cite{vacuumon}, with
early-epoch relativistic matter present, with EoS $w_m =1/3$.
{\underline \it Dashed line} : String-inspired RVM~\cite{bms1,bms2}, with
early-epoch massless `stiff' stringy (gravitational axion) ``matter'' present, with EoS $w_m =1$.
 The potentials are defined for positive values of the vacuumon
field $\phi > 0$.}}
\label{fig:pot}
\end{figure}

As explained in \cite{vacuumon}, the potentials  \eqref{Pott} or \eqref{Pott_string}
are defined only for positive values of the vacuumon field, $\phi > 0$ ($+$ branch of solutions for $\phi$ in \eqref{ff3}). They both have a
``hill-top'' shape (\joantext{see fig.~\ref{fig:pot}}),
exhibiting a local maximum for a zero value of the classical vacuumon field $\phi$, and then decaying to zero for large values, which represents the decay of the `false' vacuum representing the graceful exit from the inflationary phase \` al RVM~\cite{rvmpheno,vacuumon}. The features between the generic and string-inspired RVM are thus qulatitatively similar, with only minor, unimportant, differences in the range of the parameters.

It was stressed in \cite{vacuumon} that the
early vacuumon field is {\it classical}, hence there is no issue in attempting to
considering the effects of quantum fluctuations of $\phi$ on the ``hill-top'' potential \eqref{Pott}.
In fact, if $\phi$ were a fully fledged quantum field, such as the
conventional inflation (which is not the case here), the potential
\eqref{Pott}
provides slow-roll parameters which
fit at $2\sigma$ level the optimal range indicated by the Planck
cosmological data on single-field inflation~\cite{planck}.

It is important to note~\cite{vacuumon} at this stage that the vacuumon representation of the RVM is {\it not} smooth between early and late eras of the Universe, in the sense that at late eras one uses another configuration of the classical vacuumon field, corresponding to a late-era potential, which has a minimun at zero, for zero values of the late-era vacuumon. This non-smooth evolution is not a cause for alarm, since, as we have already mentioned, the string-inspired
RVM is also known not to exhibit a smooth evolution, due to intermediate cosmic phase transitions.
For our purposes in this article we shall only be interested in the early-era vacuumon field representation of the RVM.

\nt{Nontheless, for completeness, we would like to review briefly at this stage, the results of \cite{bms1,bms2} on how the string-inspired RVM universe evolves from the early inflationary phase till the present, dark-energy dominated era, stressing differences, as compared to the standard RVM evolution, particularly in the early universe\,\cite{rvmhistory,JSPRev2013,JSPRev2015}. }

\nt{In our scenario, the string-inspired Universe, may be characterised by a pre-inflationary phase, immediately after the `Big-Bang', dominated by purely stringy quantum gravity effects. Such an era cannot be described by a local
low-energy effective field theory, due to an (infinity) of higher-curvature and higher-derivative terms of gravitational degrees of freedom, as well as stringy states of transplanckian masses. In fact such effects might also smoothen out the initial singularity (see, e.g. the worlk of \cite{art} on how string-inspired dilaton-Gauss-Bonnet terms, quadratic in space-time curvature, alone, suffice for such a smooth behaviour at the Big-Bang). This latter feature would be in qualitative agreement with the effective RVM evolution after the aforementioned pre-inflationary phase, see e.g.  Eq\.\,\eqref{stringHdS}, which implies  finite (non-singular) values for the Hubble rate and  energy density emerging from the effective RVM form \eqref{rLRVM}.}

\nt{Such stringy-dominated eras of the Universe, may well be featured by dynamical supergravity breaking, at scales above the RVM inflationary scale, through either simple gravitino condensation~\cite{rvmsugra} or, more complicated processes, involving {\it e.g.} gaugino condensation in, say, hidden sectors of the string-inspired model~\cite{msfuture},\footnote{\nt{We remind the reader~\cite{bms1} that in our scenario only degrees of freedom are assumed present as external lines in Feynman graphs in the four dimensjonal early Universe. All other degrees of freedom such as gauge, appear as virtual quantum fluctuations, or in hidden sectors of the string-inspired model.}} {scenarios which lead to unstable domain wall networks in  the stringy Universe, which percolate, and collapse in a non-spherically symmetric way, leading to the formation of primordial gravitational waves (GW) and other metric (tensor) fluctuations.} \nt{Such perturbations might also be produced by merging of primordial black holes, formed, {\it e.g.} by massive stringy compactified brane defects that might be present in scenarios where our Universe is viewed as an uncompactified brane world~\cite{msfuture}.} The ultra-massive gravitinos (of mass higher than the string-inspired RVM inflation scale), along with the other supermassive (transplanckian) string states decouple from the effective field theory, as the cosmic time elapses, which thus is well described by \eqref{sea4}, comprising of only the massless degrees of freedom of the bosonic gravitational multiplet of the string in the broken supergravity phase.}

\nt{Condensation of GW then leads to the string-inspired RVM phase of dynamical inflation, described in \cite{bms1}. As shown in detail in \cite{bms1}, the scale of this inflation is given by the (dominant) $H^4$ term in the total RVM-type energy density \eqref{toten}, which is of order
\begin{align}
\rho_{\rm total}^{\rm inflation}  &\simeq 3\kappa^{-4} \, \Big[\frac{\sqrt{2}}{3} \, \Big[\overline b(0) \, \kappa  + \sqrt{2\, \epsilon} \,  \mathcal N \Big]  \, \times {5.86\, \times} \, 10^6 \, \left(\kappa\, H \right)^4 \Big] \nonumber \\
&\simeq 3\kappa^{-4} \, \Big[\frac{\sqrt{2}}{3} \, |\overline b(0)| \, \kappa \, \times {5.86\, \times} \, 10^6 \, \left(\kappa\, H \right)^4 \Big] > 0~.
\end{align}
for $H $ constant.
In the first line we have given the maximum order of fluctuation of the linear-varying with cosmic time gravitational KR axion background $\overline b(t) \sim H_I \, t$  during inflation, which is assumed to last $\mathcal N$ e-foldings. Such fluctuations are negligible for $\overline b(0) \gtrsim 10 \, \kappa^{-1}$, which we assumed in \cite{bms1}, and thus to leading order one obtains the de-Sitter approximation \eqref{toten}. In terms of the standard RVM \eqref{rLRVM},
the scale of inflation is determined by setting $H \simeq  H_I/\sqrt{\alpha}$ ({\it cf.} \eqref{HS1}).}

\nt{We also remark that the boundary condition $\overline b(0)$ cannot be determined within our effective field theory approach. This would require knoweledge of the underlying microscopic string theory, which, according to our previous discussion, leads to to the RVM-type inflation via the condensation of the GW. The order of the fluctuations $\sqrt{2\epsilon} \, \mathcal N \lesssim1 $ is compatible with the current accuracy of the slow-roll cosmological data via the order of the phenomenological parameter $\epsilon \sim 10^{-2}$. This itself depends on the microscopic string physics, which goes beyond the scope of our current discussion. In this respect the situation is rather similar to what characterises the case of the Starobinsky model of dynamical inflation~\cite{staro}, although, as we repeatedly stressed in \cite{vacuumon,bms1}, our physics is different from that model. }

\nt{We would also like to mention that, from  a generic RVM effective gravitational field theory point of view, primordial fluctuations can be understood, like in the Starobinsky model, as arising from higher curvature $R^2$ terms in the action.  The terms that are generated from the functional differentiation of $R^2$ in the action turn out to vanish for   $H=$const, see e.g. \cite{JSPRev2015, Cristian2020}.   The inflationary process in this case is driven by a  short period of $\dot{H}=$const.  rather than a corresponding period of $H=$const.    For this reason the presence of terms of the form $H^4$  (which cannot appear in Starobinsky inflation)  are genuine ingredients of the new underlying mechanism of inflation, namely RVM inflation,  which is characterized by the vacuum energy density  \eqref{rLRVM}.  Such terms, which as mentioned above, are present in the stringy version of the RVM model, are also part of any effective gravitational quantum field theory of RVM~\cite{Cristian2020},  and play a significant role in the early universe. In string theory models, of course, as we discussed above, such fluctuations might be realised in more microscopic mechanisms involving supersymmetry/supergravity breaking, which also proceed via higher curvature terms in the supergravity actions and can also be cast in an RVM form, as discussed in \cite{rvmsugra}.}

\nt{During the end of our inflationary era, relativistic chiral matter is generated~\cite{bms1}, as a result of the
decay of the vacuum, which are held responsible for a {\it cancellation} of the gravitational anomalies,
and so the universe enters in a post-inflationary relativistic matter dominated era, which succeeds the stiff-matter/inflationary era.
During inflation, of course, any massive GW source, including primordial black holes, is washed out. However, as stressed in \cite{bms1,bms2}, the linear in cosmic time KR axion background, generated by the GW condensate, remains undiluted and carries over to the radiation phase, leading to leptogenesis, as per the scenario of \cite{bmsa}. The important thing to notice, is that during the radiation- and matter-dominated eras, there are contributions from cosmic radiation fields, which yield a positive $\nu >0$ coefficient, in contrast to the negative $\nu <0$ during the inflationary phase of our string-inspired RVM. During the radiation/matter eras, non-perturbative (eg QCD/gluon effects) can also generate a potential for the KR field, which, thus, can play the r\^ole of a Dark Matter component~\cite{bms2}.}

\nt{A final issue we would like to mention is that in such a string universe, in the current era, at which matter has been diluted significantly, gravitational wave can again condense, leading to a current-epoch cosmological constant (approximately), via a similar mechanism to the one that lead to inflation, whose scale though is determined by the current era Hubble parameter $H_0$ which is much smaller
than the inflationary-epoch Hubble parameter. In this respect, for our string model, an evolution of the form \eqref{HS1} with $a \sim 1$ is in operation, but with the current-epoch inflationary scale $H_I/\sqrt{\alpha} \sim H_0$.}

\section{String-Inspired RVM and the Swampland} \label{sec:stringSC}

\begin{figure}
\includegraphics[width=0.65\textwidth]{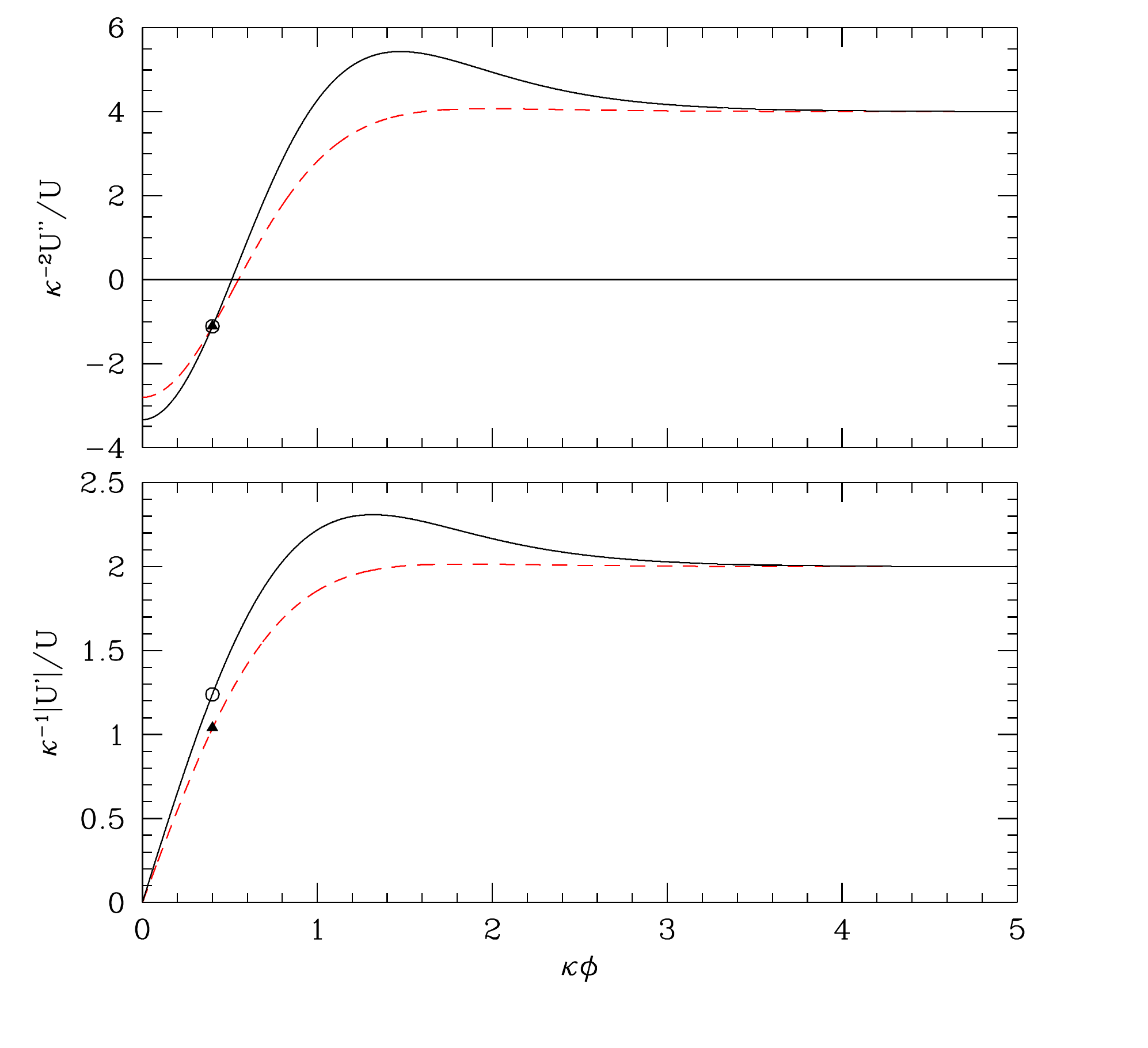}
\caption{{\it Checking the swampland criteria for the early-era RVM Vacuumon
Effective potentials, \eqref{Pott}
({\it solid} lines) for generic RVM with relativistic
matter, and \eqref{Pott_string} ({\it dashed} lines)
for string-inspired RVM with stiff stringy
(gravitational axion) ``matter''.
\underline{{\it Upper panel}}~: check of the Second Swampland Conjecture, which is
found to be satisfied (i.e. $\kappa^{-2}\, |V''|/V < -\mathcal O(1)$, near the local maximum of the potential (at $\phi =0$) for small ranges of the vacuumon field $\kappa \, \phi < 0.4$.
\underline{{\it Lower panel}}~: check of the range of validity, in
vacuumon field space, of the first Swampland conjectrure, which is found
to be satisfied ($\kappa^{-1}\, |V'|/V > 1.24$ (open circle, for generic RVM), or 1.04 (triangle, for string-inspired RVM), for large field values $\kappa \phi \gtrsim 0.4$. The asymptotic value $2$ of $|V'|/V $ for large $\kappa \phi > 1$ can be understood by the saturation of the entropy of the string-inspired RVM~\cite{bms1,bms2} by the Bousso entropy bound~\cite{Bousso} (\joantext {maximum Bekenstein}-Gibbons-Hawking entropy~\cite{Bekenstein,Hawking,GH}) during the exit from the early de Sitter phase.}}
\label{fig:SCs}
\end{figure}

The embedding of the RVM in string theory, which thus serves as the UV complete theory of the effective gravitational action \eqref{sea4}, calls for a
discussion on its compatibility with the Swampland Conjectures, which would shed light in
the nature of the UV completion of such EFT.
Despite the fact that, within the RVM framework, the inflation is not due to a fundamental scalar mode, nonetheless
the RVM contains scalar modes, such as the KR axion, as well as the scalar mode hidden in the GW condensates~\cite{bms1,bms2}, which increase with the cosmic time, and thus they constitute potentially dangerous
agents for a breakdown of the corresponding EFT, through towers of string states that may become light, as the
aforementioned scalar modes ``distance in theory space'', from an initial point, becomes large.

In this respect, we would like to reconsider our thesis~\cite{vacuumon,bms1} that during the early de Sitter phase, the vacuumon representation of the RVM is purely classical, without quantum fluctuations, and ask the question as to whether the EFT vacuumon potential \eqref{Pott}, which represents, for a given range of values of the $\phi$ field, the RVM \eqref{rLRVM}, finds a consistent  interpretation as a low-energy QFT compatible with the UV completion of the string-inspired RVM, {\it i.e.} whether it satisfies (some of) the Swampland criteria, examined in section \ref{sec:swamp}.

If this question is answered in the affirmative, it would be a first, but rather radical, step towards a demonstration of the consistency of the low-energy string-inspired approach of \cite{bms1,bms2}, as an EFT that could admit a proper quantum-gravity/string theory UV completion. This would strengthen its conclusions, and with it, those of the RVM approach, which would appear as a consistent UV complete EFT.

Let us first discuss some properties of the early vacuumon potentials \eqref{Pott} and \eqref{Pott_string}  (see fig.~\ref{fig:pot}), with respect to the Swampland Criteria \eqref{SC1a} and \eqref{SC2}. The vacuumon representation EFT involves a single real scalar field $\phi$, which grows with the cosmic time. The potentials have a local maximum at field value zero, and as such it is easy to see that they satisfy the SSC \eqref{SC2}, for small field values near this local maximum. Indeed, we observe the following:

\begin{itemize}

\item{(i)} On plotting the quantity $\kappa^{-2}\, V^{\prime\prime}/V$ versus the field $\phi >0$, for positive values, as required by the vacuumon representation~\cite{vacuumon}, we observe from fig.~\ref{fig:SCs} (upper panels)
that the SSC, \eqref{SC2}, is satisfied ($\kappa^{-2}\, V^{\prime\prime}/V \lesssim -1$)  for small vacuumon
field values  $0 < \kappa \, \phi \lesssim 0.4$ in both cases  \eqref{Pott} and \eqref{Pott_string}. This range defines the bulk of the de Sitter (inflationary) phase for which $D \, a^4 \ll 1$ or $D_{\rm string} \, a^6 \ll 1$, in the generic and string-inspired RVM cases, respectively. We interpret this result as implying
that in this range the vacuumon EFT
are consistent with an UV completion from string theory.

\item{(ii)} As the early-era vacuumon field continues to grow, exceeding a `critical range' range $\kappa \phi > 0.4$ (see discussion in section \ref{sec:swamp}), towers of string states become light and affect the EFT (, the `critical range' . This phase characterises the exit from inflation phase of the RVM, in which most of its entropy is generated due to such states. For $\kappa\, \phi \, \gg \, 1$ we observe (fig.~\ref{fig:SCs}, lower panels) that  the quantities $| V^\prime | / V$  approximate a constant value, $-2$ for the generic RVM with radiation-type matter, and $-2.45$ for the string-inspired RVM with `stiff' stringy matter.

These values coincide with the value of $V^\prime/V$ in the regime
\be\label{exit}
D \, a^4 \gg 1,  \qquad
D_{\rm string} \, a^6 \gg 1
\ee
that is assumed to characterise the exit of inflation, corresponding to large vacuumon fields $\kappa \, \phi \gg 1$.

Indeed, in this range, the entropy of the RVM,
which is mainly due to the tower of string states, that are becoming light ({\it cf.} \eqref{entropy}),
saturates the Bousso upper bound \eqref{GH} and becomes the \joantext{Bekenstein}-Gibbons-Hawking entropy of the RVM~\cite{Bekenstein,Hawking,GH}. This is a standard property of the
RVM~\cite{solaentropy}, and we used it in \cite{bms1,bms2} in order to connect smoothly the exit from inflation, corresponding to an almost constant (slowly varying) background axion field $b$, $\dot b \, \simeq \, {\rm constant}$, with a temperature dependent configuration for this field $b(T)$ during the radiation era, which leads to leptogenesis~\cite{bmsa}.

Let us now demonstrate this. Consider the saturated from below \eqref{eSC1b} and saturate {\it also} the Bousso bound for the entropy \eqref{GH},
assuming that the tower of string states that descend from the UV are mainly {\it point like}, which implies $\delta \simeq 0$. We have then:
\be\label{eSC1b2}
\frac{|V^\prime |}{V} \simeq (\ln [N^\gamma])^\prime = -2 \, \frac{H^\prime}{H}, \quad \kappa\, \phi \gg 1.
\ee
We now use  the set of equations \eqref{HS1rad}  together with \eqref{Pott}, or \eqref{stringHdS} together with \eqref{Pott_string},
depending on whether we consider generic RVM with relativistic matter~\cite{rvmhistory} or the string-inspired RVM with stiff stringy $b$-axion matter of \cite{bms1,bms2}, respectively. We consider the range \eqref{exit}, appropriate for the exit-from-inflation phase.
 From \eqref{eSC1b2},  we then obtain
\be\label{eSC1b22}
\frac{|V^\prime |}{V} \simeq (\ln [N^\gamma])^\prime = -2 \, \frac{H^\prime}{H}
\ee
which yields
\be\label{eSC1b2a}
\frac{|V^\prime |}{V} \simeq  4 \, \frac{1}{a}\, \frac{da}{d\phi}  \simeq  2\, \kappa\, , \quad D\, a^4 \gg 1,
\ee
for the generic RVM, and
\be\label{eSC1b2b}
\frac{|V^\prime |}{V} \simeq  6 \, \frac{1}{a}\, \frac{da}{d\phi}  \simeq  \sqrt{6} \, \kappa\, \simeq 2.45 \, \kappa\,, \quad D_{\rm string}\, a^4 \gg 1
\ee
for the string-inspired RVM.

The reader can readily verify from the lower panels of fig.~\ref{fig:SCs} that these are indeed the asymptotic limit
of the function $\frac{|V^\prime |}{V}$ for large $\kappa \, \phi$.

\end{itemize}
Thus, the two cases, RVM with relativistic matter, and string-inspired RVM with stiff stringy matter,
exhibit very similar behaviour in order of magnitude and they can both be embedded in an UV complete string theory model.

In RVM, there is interaction between the `running vacuum' and `matter', as per equation \eqref{frie33}, but, as we have already mentioned, during the de Sitter phase, the right-hand-side of this equation is expected to be strongly subleading, of order $(q+1)H^5$ and higher, hence, as in the Warm Inflation case of \cite{branden}, one expects any potential modifications to the SSC, as a result of such an interaction, to be very small.

\section{Conclusions, Vacuumon and the Weak Gravity Conjecture}\,\label{sec:conclusions}

We have argued in favour of the vacuumon representation of RVM satisfying the Second Swampland Criterion \eqref{SC2} for appropriate range of the vacuumon fields. This is due to the fact that the vacuumon effective potentials  \eqref{Pott} or \eqref{Pott_string} are sufficiently ``tachyonic'' near the origin in field space. As such, they do not satisfy the slow roll condition for inflation, and in this sense they cannot be used to fit phenomenological data~\cite{planck}.

In the context of the string-inspired RVM~\cite{bms1,bms2}, it is the gravitational axion $b(x)$ field that exhibits a slow roll behaviour during the inflationary era, which is induced by gravitational-wave condensates.
Although the initial value $b(0)$ of this field is transplanckian, nonetheless the effective action depends only on the derivatives of this field, which satisfy the slow roll condition $\dot b \sim 10^{-2} \, H\, \kappa^{-1}  $, with $H \sim H_I \sim 10^{-4} \MPl$, and so is sub-Planckian, not presenting a problem for the EFT.

The above analysis demonstrated that the small-field behaviour of the vacuumon potential is compatible
with an UV complete theory, such as strings, and so one might view this effective description under a new scope.
In fact, the vacuumon provides a faithful description  of the decay of the RVM and the passage from the
inflationary to radiation dominated era, in the context of the string-inspired model of \nicktext{\cite{bms1,bms2,bms3}}. The satisfaction of the second swampland conjecture for small vacuumon fields, and the saturation of the Bousso entropy bound at the end of the inflationary period, associated with transplanckian values of the vacuumon, as explained above, implies a potential embedding of the model to UV complete microscopic string theory models in a consistent way. From this point of view, the upshot of the above note is to point out that it is the tower of massless string states that are becoming light at the end of inflation, and thus contaminate the effective gravitational theory of the early de Sitter era of \cite{bms1,bms2}, which are responsible for the emergence of relativistic matter/radiation at the end of inflation, leading to a radiation dominated Universe after the exit from inflation. \joantext{Such  a massive pouring of relativistic states from the string  into the effective field theory domain, could be the kind of graphical picture provided by the swampland criteria in order to explain the huge generation of entropy in the initial stages of  RVM-inflation.}

\begin{figure}
\includegraphics[width=0.6\textwidth]{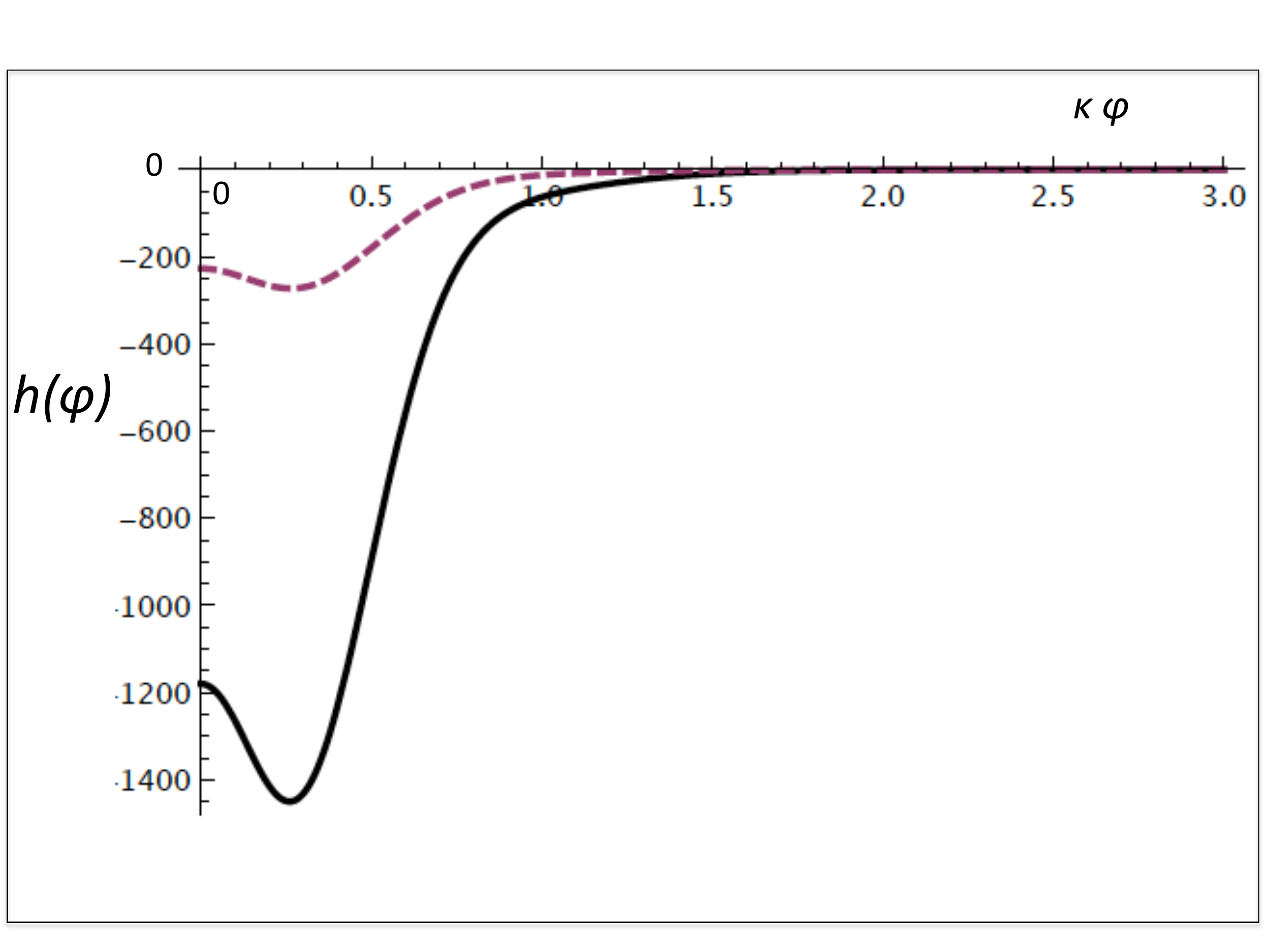}
\caption{{\it Checking the Weak-Gravity Conjecture (WGC) for scalars, as formulated in \cite{ibanezWGC}, \eqref{WGC}, for the early-era RVM Vacuumon Effective potentials. The quantify $h(\phi)= \kappa^2 (V^{\prime\prime})^{2}-2 (V^{\prime\prime\prime})^2+V^{\prime\prime}V^{\prime\prime\prime\prime}$ is plotted as a function of
$\kappa \phi$, for the cases of: (i) the potential  \eqref{Pott} (continuous line), corresponding to
 the Generic RVM~\cite{vacuumon}, with
early-epoch relativistic matter present, with EoS $w_m =1/3$;
(ii) the potential \eqref{Pott_string} (dashed line), corresponding to the String-inspired RVM~\cite{bms1,bms2}, with early-epoch massless `stiff' stringy (gravitational axion) ``matter'' present, with EoS $w_m =1$. We work in units with $V_{0}={\tilde V}_{0}=1$ and $\kappa=1$. The quantity $h(\phi)$ remains negative for the entire range of the vacuumon field $\kappa \phi > 0$, which implies the satisfaction of
  \eqref{WGC}.}}
\label{fig:WGC}
\end{figure}

Part of this matter, pertains to chiral fermions, which are responsible for the cancellation of the primordial gravitational anomalies, as discussed in \cite{bms1,bms2}. In the notes above we have assumed that these extra (light) string states behave {\it mostly} as point particles, which allowed us to take $\delta \simeq 0$ when estimating their entropy \eqref{entropy}. This gave consistent results with the vacuumon picture, as we explained above. Of course, as string/brane theory is characterised by a landscape of mathematically consistent models, one may consider also other situations in which cosmic strings and membranes also descend from the UV. The results  of the previous section do not change much if one uses a generic
$2 > \delta > 0$, to take into account these more general cases. Of course the limiting case of $\delta=2$ is problematic, as this would lead to a not well defined swampland criterion \eqref{eSC1b}, unless the parameter $\gamma$ approaches  zero at least as fast as $\delta \to 2$. If the approach of $\gamma $ to zero is faster than the approach of $\delta$ to 2, then the swampland criterion becomes trivial, otherwise the bound is not determined. However, one could argue that such a limiting case would correspond to the unphysical situation  where (two-spatial dimensional) domain walls dominate the exit from inflation, which do not consider here.

\joantext{All that said, the  failure in reproducing the slow-roll behavior of the RVM through the vacuumon picture can be conceived as a counter-example against ascribing direct  physical meaning  to a scalar field representation fulfilling the Swampland criteria.  There is,  in general, no one-to-one correspondence between the scalar field representation and the original theory. The Swampland criteria intend to put forward a set of conditions, which, if satisfied, do not necessarily imply anything definite on the effective field theory, but if not satisfied, then the model may be in trouble. In our case they are satisfied, so the model may be thought of as a consistent quantum gravity model, but the reality is that it is not the full fledged quantum description of the RVM as derived from the low-energy effective action of  the graviton and antisymmetric tensor (spin-one) KR fields of the massless (bosonic) string gravitational multiplet. Even so, the nice interpretation of the onset of the radiation-dominated epoch as caused from the large transfer of massive states exponentially becoming ultralight (hence relativistic) owing to the properties of the vacuumon field, somehow points to a nice complementary image of the RVM universe, and provides an additional connection to its string-inspired interpretation.}

A final comment concerns the satisfaction, by our vacuumon potentials, of the (appropriate) version of the so-called {\it Weak Gravity Conjecture} (WGC) for scalars, as formulated in~\cite{ibanezWGC}, Eq. \eqref{WGC}.  It is straightforward to see (see fig,~\ref{fig:WGC}), that our vacuumon potentials \eqref{Pott}, \eqref{Pott_string} satisfy this condition for the entire range of the (positive) vacuumon fields $\phi > 0$. This strengthens our view that the vacuumon might indeed be more than a classical field description of the RVM, and might contain the seeds for a full quantum scalar mode description.

There are of course many issues that  remain open in the model of \cite{bms1,bms2}, such as a proper treatment of the gravitational-wave condensate and a study of its effective potential, which will allow for a complete understanding of the connection of the RVM inflation with a true scalar mode. Also a further understanding of the phenomenoplogy of the de Sitter phase of this string-inspired RVM model, in  the sense of identifying `smoking-gun' signatures, if any, of its stringy nature, for example evidence in the
early Universe data of the negative $\nu$ coefficient of the $H^2$ term in the energy density \eqref{rLRVM}, which appears to characterise the de Sitter phase of the stringy RVM model.
Despite such open issues, we believe that the results discussed here offer support for the UV consistency of the string-inspired RVM model of \cite{bms1,bms2},
encouraging further studies. We hope to be able to tackle some of the above issues in future works.

{\it Affaire \`a suivre...}

\section*{Acknowledgements}

The work  of NEM is supported in part by the UK Science and Technology Facilities  research Council (STFC) under the research grants
ST/P000258/1 and ST/T000759/1. The work of JS has
been partially supported by projects  FPA2016-76005-C2-1-P (MINECO), 2017-SGR-929 (Generalitat de Catalunya) and MDM-2014-0369 (ICCUB). SB acknowledges support from
the Research Center for Astronomy of the Academy of Athens in the
context of the program  ``{\it Tracing the Cosmic Acceleration}''.
This work is also partially supported by the COST Association Action CA18108 ``{\it Quantum Gravity Phenomenology in the Multimessenger Approach (QG-MM)}''.
NEM acknowledges a scientific associateship (``\emph{Doctor Vinculado}'') at IFIC-CSIC-Valencia University, Valencia, Spain.

\end{document}